\documentstyle[12pt]{article}
\newcounter{dummy}{}
\newcommand{\letters}{\setcounter{dummy}{\value{equation}}
\refstepcounter{dummy}
\setcounter{equation}{0}
\renewcommand{\theequation}{\arabic{dummy}\alph{equation}}}
\newcommand{\noletters}{\setcounter{equation}{\value{dummy}}
\renewcommand{\theequation}{\arabic{equation}}}
\newenvironment{mathletters}{\letters}{\noletters}
\date{}
\title{Instantons in non-Cartesian coordinates}
\author{A.~A.~Abrikosov,~jr. %
\thanks{The work was done with partial support of the RFBR grant
00-02-17808.} \\
{\em ITEP, 117218 Moscow, RUSSIA}}
\begin{document}
\maketitle

\begin{abstract}
The explicit multi-instanton solutions by 'tHooft and Jackiw, Nohl \&
Rebbi are generalized to curvilinear coordinates. The idea is that a
gauge transformation can notably simplify the expressions obtained after
the change of variables.  The gauge transform generates a compensating
addition to the gauge potential of pseudoparticles.  Singularities of the
compensating field are irrelevant for physics but may affect gauge
dependent quantities.
\end{abstract}

\section*{Introduction}

The years that passed since the discovery of instantons \cite{BPST} did
not bring the answer to the question about the role of instantons in QCD
\cite{Shifman,Schaefer/Shuryak}.  As far as confinement remains a puzzle
all references to instantons at long scales are ambiguous. Indications may
come from studies of instanton effects in phenomenological models.  These
could tell whether confinement may seriously affect pseudoparticles and
{\em v.~v.}

Common confinement models look most natural in non-Cartesian coordinate
frames. The obvious choice for (spherical) bags are 3+1-cylindrical, {\em
i.~e.\/} 3-sphe\-ri\-cal+time, coordinates while strings would prefer
2+2-cy\-lin\-dri\-cal (2+1-cy\-lin\-dri\-cal+time) geometry. Nevertheless
up till now instantons were usually discussed in the Cartesian frame (that
was ideal in vacuum). In the present work we shall generalize to
curvilinear coordinates the multi-instanton solutions by 'tHooft and
Jackiw, Nohl \& Rebbi \cite{JNR} and simplify the formulae by the gauge
transformation. I would expect that the procedure makes sense for the AHDM
\cite{AHDM} solution and other topological configurations as well.

We shall start from reminding the basics of curvilinear coordinates in
Sect.~{\ref{cc}} where the Levi-Civita connection and the spin connection
are described. In Sect.~{\ref{inst}} we introduce the multi-instanton
solutions known explicitly.  In Sect.~{\ref{mult-inst}} we shall rewrite
these formulae in non-Cartesian coordinates and propose the gauge transform
that makes them compact. The price will be the appearance of the
addition to the original field that we shall call the {\em
compensating\/} gauge connection. The example of the $O(4)$-sphe\-ri\-cal
coordinates is sketched in Sect.~{\ref{examp}}.  Singularities of the
gauged solution are discussed in Sect.~{\ref{sing}}.  The last part
summarizes the results.

\section{Basics}   \label{basics}

\subsection{Curvilinear coordinates}  \label{cc}

We shall consider flat 4-dimensional euclidean space-time that may be
pa\-ra\-met\-rized either by a set of Cartesian coordinates $x^\mu $ or
by curvilinear ones called $q^\alpha $. The $q$-frame is characterized by
the metric tensor $g_{\alpha \beta }(q)$ so that:
\begin{equation}
ds^2=dx_\mu ^2=g_{\alpha \beta }(q)\,dq^\alpha \,dq^\beta .
\end{equation}

Now the partial derivatives $\partial_\mu = \frac\partial{\partial x^\mu}$
must be replaced by the covariant ones $D_\alpha $. For example the
derivative of a contravariant vector $A^\beta $ is:
\begin{equation}
D_\alpha A^\beta =\partial _\alpha A^\beta
+\Gamma _{\alpha \gamma}^\beta \, A^\gamma .
\end{equation}
The function $\Gamma _{\beta \gamma}^\alpha$ is called the
{\bf Levi-Civita connection.} It can be expressed in terms of the metric
tensor and its inverse $g^{\alpha \beta}$:
\begin{equation}
\Gamma _{\beta \gamma }^\alpha =
\frac 12 g^{\alpha \delta }
\left( \frac{\partial g_{\delta \beta }}{\partial q^\gamma }
+\frac{\partial g_{\delta \gamma }}{\partial q^\beta }
-\frac{\partial g_{\beta \gamma }}{\partial q^\delta }\right) .
\label{Gamma}
\end{equation}

Often it is helpful to use instead of $g_{\alpha \beta }$ the set of four
vectors $e_\alpha ^a$ called the vierbein:
\begin{equation}
g_{\alpha \beta }(q)=\delta_{ab} \, e_\alpha ^a(q) \,e^b_\beta (q).
\label{g<=>pi}
\end{equation}
Multiplication by $e_\alpha ^a$ converts coordinate (Greek) indices into the
vierbein (Latin) ones,
\begin{equation}
A^a=e_\alpha ^a\,A^\alpha .
\end{equation}

Covariant derivatives of quantities with Latin indices are defined in terms
of the {\bf spin connection }$R_{\alpha \,b}^a(q)$,
\begin{equation}
D_\alpha A^a=\partial _\alpha A^a+R_{\alpha \,b}^a\,A^b.
\end{equation}

The two connections $\Gamma _{\alpha \delta }^\beta $ and $R_{\alpha \,b}^a$
are related to each other as follows (note that both $g_{\alpha\beta}$
and $e^a_\alpha$ are covariantly constant by construction):
\begin{equation}
R_{\alpha \,b}^a=e_\beta ^a\,\partial _\alpha e_b^\beta +e_\beta ^a\,\Gamma
_{\alpha \gamma }^\beta \,e_b^\gamma =e_\beta ^a\,(D_\alpha e^\beta )_b.
\label{R_alpha}
\end{equation}

Decomposition (\ref{g<=>pi}) of the metric into vierbeins is not unique.
It is defined up to orthogonal transformations of the vierbein with
respect either to the Greek or to Latin indices. Two vierbeins of the same
orientation (that means those with the same signs of
$\det ||e^a_\alpha||$) are related by a simple $O(4)$ rotation.
Orientation of the vierbein can be changed by reversing one of its
components, say, $e^a_1 \rightarrow -e^a_1$.

It is convenient to use the freedom in order to make some components of
vierbein zero. In general $e^a_\alpha$ is a $4\times 4$-matrix.
However if the coordinate frame is orthogonal%
\footnote{%
This class incorporates many coordinate systems including spherical,
cylindric, parabolic, elliptical and others.}
and the metric tensor may be diagonalized,
\begin{equation}
g_{\alpha \beta}(q) = G(q)\, \delta_{\alpha \beta},
\label{g_diag}
\end{equation}
then one can use the diagonal vierbein
\begin{equation}
e^a_\alpha = \sqrt{G(q)} \delta^a_\alpha.
\label{e_diag}
\end{equation}
We shall call it the {\em natural\/} vierbein. Only four of its components
are not zero.  From here on we shall assume that the curvilinear system is
orthogonal so that the natural vierbein exists.

\subsection{Instantons}    \label{inst}

We shall discuss the pure euclidean Yang-Mills theory with the $SU(2)$
gauge group. The vector potential is
$\hat{A}_\mu = \frac 12 \tau^a\, A_\mu^a$ where $\tau ^a$ are the Pauli
matrices. The (Cartesian) covariant derivative in fundamental
representation is $D_\mu =\partial _\mu -i\,\hat{A}_\mu $, and the action
has the form:
\begin{equation}
S=\int \frac{{\rm tr}\, \hat{F}_{\mu \nu }^2}{2e^2}\,d^4x=
\int \frac{{\rm tr}\,\hat{F}_{\alpha \beta }\,
\hat{F}^{\alpha \beta }}{2e^2}\,\sqrt{g} \, d^4q.
\label{s_gauge}
\end{equation}
where $g=\det \left| \left| g_{\alpha \beta }\right| \right|$ and $e$ is
the coupling constant. The formula for the gauge field strength
$\hat{F}_{\alpha \beta }$ is universal:

\begin{equation}
\hat{F}_{\alpha \beta }(\hat{A})=\partial _\alpha \hat{A}_\beta -\partial
_\beta \hat{A}_\alpha -i\,\left[ \hat{A}_\alpha ,\,\hat{A}_\beta \right] .
\label{[DmuDnu]}
\end{equation}
The action is invariant under gauge transformations
\begin{equation}
\hat{A}_\mu \rightarrow \hat{A}_\mu ^\Omega =\Omega ^{\dagger }\,\hat{A}_\mu
(x)\,\Omega +i\,\Omega ^{\dagger }\,\partial _\mu \Omega \,,
\label{A_s=OA_rO+OdO}
\end{equation}
where $\Omega $ is a unitary $2\times 2$ matrix, $\Omega ^{\dagger }=\Omega
^{-1}.$

The field equations have selfdual ($F_{\mu \nu }=\tilde{F}_{\mu \nu
}=\frac 12\epsilon _{\mu \nu \lambda \sigma }\,F^{\lambda \sigma }$)
solutions known as instantons. It happens that all explicit multi-instanton
solutions known by now are described by the same formula:
\begin{equation}
\hat{A}_\mu (x)=-\frac{\hat{\eta}_{\mu \nu }^{-}}2 \,
\partial_\nu \ln \rho (x),
\label{JNR}
\end{equation}
where $\hat{\eta}_{\mu \nu }$ is the matrix version of the 'tHooft's
$\eta$-sym\-bol, \cite{tHooft}:
\begin{equation}
\hat{\eta}_{\mu \nu }^{\pm }=-\hat{\eta}_{\nu \mu }^{\pm }=\left\{
\begin{array}{cc}
\tau ^{a\,}\epsilon ^{a\mu \nu }; & \mu ,\nu =1,2,3; \\
\pm \tau ^a\,\delta ^{\mu a}; & \nu =4.
\end{array}
\right.   \label{eta^pm}
\end{equation}

Depending on the choice of the scalar function $\rho(x)$ expression
(\ref{JNR}) describes either the higher selfdual configurations found by
Jackiw, Nohl and Rebbi and 'tHooft's {\em Ansatz\/}  \cite{JNR} or
instantons in singular and regular gauges (the latter requires the
substitution $\hat \eta^- \rightarrow \hat \eta^+$ (\ref{A_reg})).

Our aim is to generalize the solution (\ref{JNR}) to curvilinear
coordinates. We shall not refer to any particular form of $\rho (x)$ and
the results will be applicable to all the cases.

\section{Instantons in curvilinear coordinates} \label{mult-inst}

\subsection{The question and the answer}

{\sc Problem} \qquad It is not a big deal to transform the covariant
vector $\hat{A}_\mu $ (\ref {JNR}) to $q$-coordinates. However this
makes the constant numerical tensor $\hat{\eta}_{\mu \nu }$
coordinate-dependent:
\begin{equation}
\hat{\eta}_{\mu \nu }\rightarrow \hat{\eta}_{\alpha \beta } =
\hat{\eta}_{\mu \nu } \frac{\partial x^\mu }{\partial q^\alpha }
\frac{\partial x^\nu }{\partial q^\beta } =
\hat{\xi}_{ab}\, c^a_\alpha (q) \, c^b_\beta (q).
\label{eta_x->eta_q}
\end{equation}
Here $\hat{\xi}_{ab}$ is a constant numerical matrix tensor that takes the
place of $\hat{\eta}_{\mu \nu }$ in non-Cartesian coordinates:
\begin{equation}
\hat{\xi}_{ab}=\delta _a^\mu \,\delta _b^\nu \,\hat{\eta}_{\mu \nu }.
\label{xi=eta}
\end{equation}
The former Cartesian vierbein%
\footnote{%
The advantage of this vierbein is that it nullifies the spin connection.}
$c^a_\mu = \delta^a_\mu$ becomes a rather complicated $4 \times 4$ matrix
$c^a_\alpha = \delta^a_\mu \frac{\partial x^\mu}{\partial q^\alpha}$ when
expressed in $q$-coordinates.  The question is if there is a way to
replace it by the simpler natural vierbein $e^a_\alpha$. \smallskip

{\sc solution}\qquad
Let us chose the orientation of natural vierbein to coincide with that of
$c^a_\alpha$. Then these two can be rotated into each other. The $\hat
\eta$-sym\-bols project spatial rotations onto the algebra of the $SU(2)$
gauge group. This means that rotations of the vierbein may be compensated by
appropriate gauge transformations and
\begin{equation}
\Omega ^{\dagger }\,\hat{\eta}_{\alpha \beta }\Omega
=e_\alpha ^a\,e_\beta ^b\, \hat{\xi}_{ab}.
\label{Oxi=etaO}
\end{equation}

The gauge-rotated instanton field is the sum of the two pieces
(\ref{A_s=OA_rO+OdO})
\begin{equation}
\hat{A}_\alpha ^\Omega (q)=
-\frac 12 e_\alpha ^a\,\hat{\xi}_{ab}\,e^{b\,\beta}\,
\partial _\beta \ln \rho \,(q)
+i\,\Omega ^{\dagger }\,\partial _\alpha \Omega.
\label{A^I+A^comp}
\end{equation}
The first addend is almost traditional and does not depend on the
$\Omega $-matrix. The second one is entirely of geometrical origin and
carries the information about the $q$-frame. We call it the {\bf
compensating connection} because it compensates the coordinate dependence
of $\hat{\eta}_{ab}=e_a^\alpha\, e_b^\beta\, \hat{\eta}_{\alpha\beta}$ and
reduces it to the constant $\hat{\xi}_{ab}$.

So long we did not specify the duality of $\hat{\eta}$-symbol.  However
the $\Omega $-matrices and compensating connections
$\hat{A}_\alpha ^{{\rm comp}\,\pm}$ are different for $\hat{\eta}^{+}$ and
$\hat{\eta}^{-}$. In general they are respectively the selfdual and
antiselfdual projections of the spin connection onto the gauge group:
\begin{equation}
\hat{A}_\alpha ^{{\rm comp\,}\pm }=i\,\Omega _{\pm }^{\dagger }\,\partial
_\alpha \Omega _{\pm }=-\frac 14\,R_\alpha ^{ab}\,\hat{\xi}_{ab}^{\pm }.
\label{A^comp}
\end{equation}

The last formula does not contain $\Omega $ that has dropped out of the
final result. In order to write down the multi-instanton solution one
needs only the vierbein and the associated spin connection.

\subsection{Triviality of the compensating field.}

The fact that the compensating connection $\hat{A}^{\rm comp} = -\frac
14\, R_\alpha ^{ab}\,\hat{\xi}_{ab}$ is a pure gauge (\ref{A^comp})
is specific to the flat space.  It turns out that the field strength
$\hat{F}_{\alpha \beta }(\hat{A}^{{\rm comp}})$ is related to the Riemann
curvature of the space-time $R_{\,\alpha \beta }^{\quad \gamma \delta }$:
\begin{equation}
\hat{F}_{\alpha \beta }(\hat{A}^{{\rm comp\,}\pm })
=-\frac 14 R_{\alpha
\beta }^{\quad \gamma \delta }\hat{\xi}_{\gamma \delta }^{\pm }.
\label{F=R/4}
\end{equation}
In the flat space $R_{\,\alpha \beta }^{\quad \gamma \delta } = 0$ and
consequently $\hat{F}_{\alpha \beta }^{}(\hat{A}^{{\rm comp}}) = 0$.
Simple changes of variables $x^\mu \rightarrow q^\alpha$ do not generate
curvature and $\hat{A}^{\rm comp}$ is a pure gauge.  However in curved
space-times the connection given by the last of expressions (\ref{A^comp})
may be nontrivial.

\subsection{Duality and topological charge}

As long as we limit ourselves to identical transformations the
vector potential (\ref{A^I+A^comp}) must satisfy the classical field
equations. However the duality equation looks differently in non-Cartesian
frame. If written with coordinate indices it is:
\begin{mathletters}
\begin{equation}
\hat{F}_{\alpha \beta }=
\frac{\sqrt{g}}2\,\epsilon _{\alpha \beta \gamma \delta }\,
\hat{F}^{\gamma \delta }.
\label{dual-greek}
\end{equation}
Still it retains the familiar form in the vierbein notation:
\begin{equation}
\hat{F}_{ab} =  \frac 12\, \epsilon _{abcd}\,\hat{F}^{cd}.
\label{dual-lat}
\end{equation}
\end{mathletters}
The topological charge is given by the integral
\begin{mathletters}
\label{q}
\begin{equation}
q=\frac 1{32\pi ^2} \int \epsilon _{\alpha \beta \gamma \delta }\,
{\rm tr\,}\hat{F}^{\alpha \beta }\,\hat{F}^{\gamma \delta }\,d^4q,
\label{q-greek}
\end{equation}
which in the vierbein notation becomes
\begin{equation}
q=\frac 1{32\pi ^2}\int \epsilon _{abcd}\,
{\rm tr\,}\hat{F}^{ab}\,\hat{F}^{cd}\,\sqrt{g}\,d^4q.
\label{q-lat}
\end{equation}
\end{mathletters}

The general expression for $\hat{F}_{\alpha \beta }$ in non-Car\-te\-sian
frame is rather clumsy \cite{9906008} but it simplifies for one instanton.
The Cartesian vector potential in regular gauge is, ($r^2=x_\mu ^2$):
\begin{equation}
\hat{A}_\mu ^I =
- \frac{\hat{\eta}_{\mu \nu }^+}2 \,
\partial_\nu \ln \frac{\rho^2}{r^2+\rho^2} .
\label{A_reg}
\end{equation}
The conjugated coordinate and $\Omega _{+}$ gauge transformations convert
it into
\begin{equation}
\hat{A}_\alpha ^I =
- \frac 12 e_\alpha ^a\, \hat{\xi}_{ab}\, e^{b\,\beta}\,
\partial_\beta \ln \frac{\rho^2}{r^2+\rho^2} +
\hat{A}^{{\rm comp}\,+},
\end{equation}
and the field strength becomes plainly selfdual:
\begin{equation}
\hat{F}_{ab}(\hat{A}_{}^{I\,})=
-\frac{2\,\hat{\xi}_{ab}^{+}}{\left( r^2+\rho ^2\right) ^2},
\label{F_reg}
\end{equation}
This generalizes the regular gauge to any curvilinear coordinate system.

\section{Example}  \label{examp}

We shall consider the instanton placed at the origin of the $O(4)$-spherical
coordinates. Those are the three polar angles and the radius:
$q^\alpha =(\chi ,\,\phi ,\,\theta ,\,r)$. The polar axis is aligned with
$x^1$ and
\begin{mathletters}
\begin{eqnarray}
x^1 &=&r\cos \chi ; \\
x^2 &=&r\sin \chi \sin \theta \cos \phi ; \\
x^3 &=&r\sin \chi \sin \theta \sin \phi ; \\
x^4 &=&r\sin \chi \cos \theta .
\end{eqnarray}
\end{mathletters}
The properly oriented natural vierbein for spherical coordinates is:
\begin{equation}
e_\alpha ^a  =
{\rm diag}\,(r,\,r\sin \chi \sin \theta ,\,r\sin \chi ,\,1),
\end{equation}

Now one may start from the vector potential (\ref{A_reg}) and step by step
carry out the entire procedure. But to the straightforward calculation of
the instanton part this requires computing $\Gamma_{\beta \gamma}^\alpha$,
then $R_\alpha ^{ab}$ and finally $\hat{A}^{{\rm comp}\,+}$. The result
is:
\begin{mathletters}
\label{instO(4)}
\begin{eqnarray}
\hat{A}_\chi ^I &=&
\frac{\tau _x}2\left( \frac{r^2-\rho ^2}{r^2+\rho ^2} \right) ;
\label{instO(4)a} \\
\hat{A}_\phi ^I &=&
-\frac{\tau _x}2\cos \theta +
\frac{\tau _y}2\sin \chi \sin \theta
\left( \frac{r^2-\rho ^2}{r^2+\rho ^2}\right)   \nonumber \\
&&+\frac{\tau _z}2\cos \chi \sin \theta ;
\label{instO(4)b} \\
\hat{A}_\theta ^I &=&
-\frac{\tau _y}2 \cos \chi +
\frac{\tau _z}2\sin \chi
\left( \frac{r^2-\rho ^2}{r^2+\rho ^2}\right) ;
\label{instO(4)c} \\
\hat{A}_r^I &=&0.  \label{instO(4)d}
\end{eqnarray}
\end{mathletters}
The corresponding field strength is given by (\ref{F_reg}).

\section{Singularities}    \label{sing}

Note that the vector field (\ref{instO(4)}) is singular since neither
$\hat{A}_\theta ^I$ nor $\hat{A}_\phi ^I$ go to zero at $\chi =0,\, \pi$
and $\theta =0,\, \pi$. As a result they change stepwise across the
Cartesian $x_1 x_4$-plane. This singularity is produced by the
compensating gauge transformation and must not affect observables.
However it may tell on gauge variant quantities. We shall demonstrate that
for the Chern-Simons number.

The topological charge (\ref{q}) can be represented by the surface
integral, $q=\oint K^\alpha \,dS_\alpha$
\cite{Shifman,Schaefer/Shuryak}. Here
\begin{equation}
K^\alpha =
\frac{\epsilon ^{\alpha \beta \gamma \delta }}{16\pi ^2}
\,{\rm tr\,}\left( \hat{A}_\beta \,\hat{F}_{\gamma \delta }
+\frac{2i}3\hat{A}_\beta \hat{A}_\gamma \hat{A}_\delta \right) .
\end{equation}
Even though $q$ is invariant $K^\alpha $ depends on gauge. Consider the
Cartesian instanton in the $\hat{A}_4=0$ gauge. The two contributions to the
topological charge come from the $x_4=\pm \infty $ hyperplanes,
$q= N_{\rm CS}(\infty)- N_{\rm CS}(-\infty)$, and the quantity
\begin{equation}
N_{\rm CS}(t)=\int_{x_4=t}K^4\,dS_4
\end{equation}
is  called the Chern-Simon number. Instanton is a transition between two
3-di\-men\-sional vacua with $\Delta N_{{\rm CS}}=1$.

Analysis of (\ref{instO(4)}) reveals a striking resemblance with this
case.  By coincidence here again $\hat{A}_4=0,$ (\ref{instO(4)d}). This
gives the idea to treat $r$ like a time coordinate attributing the
Chern-Simons number $N_{{\rm CS}}(r)$ to the sphere of radius $r$. A naive
expectation would be that
$\Delta N_{{\rm CS}}=\left. N_{{\rm CS}}(r)\right|_0^\infty $ gives the
topological charge. However this is not true and
$\left. N_{{\rm CS}}(r)\right|_0^\infty = \frac 12$. The second half of
$q$ is contributed by the singularities at $\theta =0,\,\pi $. We see that
the gauge transformation has affected the distribution of $N_{{\rm CS}}$.

We conclude that in our approach gauge variant quantities depend on the
choice of coordinate system and may be localized at the singularities of
the $\Omega$-transform. This may be another promising possibility to
simplify calculations with the help of curvilinear coordinates.

\section*{Summary}

We have shown how the explicit {\em (multi-)\/}instanton solutions can be
generalized to curvilinear coordinates. The gauge transformation converts
the vierbein-dependent $\hat{\eta}_{ab}$-symbol into the constant
numerical matrix $\hat{\xi}_{ab}$. The gauge potential is a sum of the
instanton part and the compensating gauge connection (\ref{A^I+A^comp}).

There is no need to know the manifest form of the gauge transform in
order to calculate the compensating connection. The computation proceeds
as follows:
\begin{enumerate}
\item  One starts from calculating the Levi-Civita connection
$\Gamma _{\beta \gamma }^\alpha $, (\ref{Gamma}).
\item  Covariant differentiation of the vierbein leads
to the spin connection $R_\alpha ^{ab}$, (\ref{R_alpha}).
\item  Convolution of the spin connection with the appropriate
$\hat{\xi}_{ab}$ gives the compensating gauge potential, (\ref{A^comp}).
\end{enumerate}

The advantage of our solution is that it is constructed directly of
geometrical quantities, {\em i.~e.\/} of the vierbein and the spin
connection. I hope that it may be useful for studies of instanton effects
in geometrically nontrivial phenomenological models. More details may be found in \cite{9906008}.

\section*{Acknowledgments}

I would like to thank the Organizing Committee for the financial support
which made possible my participation in the conference. The wonderful
choice of the location and the atmosphere of the conference also deserve
cordial compliments. I greatly appreciate the friendly and enlightening
criticism by I.~V.~Tyutin.



\begin{thebibliography}{99}
\def\journal#1#2#3#4{{#1},\ {#2}\ {(#3)}\ {#4}}
%
\def\NPB{\journal{Nucl.\ Phys.\ \mbox{\rm B}}}
\def\PLA{\journal{Phys.\ Lett.\ \mbox{\rm A}}}
\def\PLB{\journal{Phys.\ Lett.\ \mbox{\rm B}}}
\def\PRD{\journal{Phys.\ Rev.\ \mbox{\rm D}}}
\def\RMP{\journal{Rev.\ Mod.\ Phys.}}
\bibitem{BPST} A.~A.~Belavin, A.~M.~Polyakov, A.~S~Schwartz,
Yu.~S.~Tyupkin, \PLB{59}{1975}{85}.
\bibitem{Shifman} M.~A.~Shifman, (ed.) Instantons in gauge theories,
World Scientific, Singapore, 1994.
\bibitem{Schaefer/Shuryak} T.~Sch\"afer, E.~V.~Shuryak,
\RMP{70}{1998}{323}.
\bibitem{JNR} R.~Jackiw, C.~Nohl, C.~Rebbi, \PRD{15}{1977}{1642}.
\bibitem{AHDM} M.~F.~Atiah, N.~J.~Hitchin, V.~G.~Drinfeld,
Yu.~I.~Manin, \PLA{65}{1977}{185}.
\bibitem{tHooft} G.~'tHooft, \PRD{14}{1976}{3432}.
\bibitem{9906008} A.~A.~Abrikosov,~jr., to appear in Nucl. Phys.~B.
See also hep-th/9906008, (1999).

\end{thebibliography}
\end{document}